\documentclass[12pt]{iopart}

\usepackage{iopams}  
\usepackage{cite}
\usepackage{cleveref}

\newtheorem{proposition}{Proposition}
\newtheorem{corollary}{Corollary}

\begin{document}

\title[SUSY versions of the Fokas--Gel'fand formula for immersion]{Supersymmetric versions of the Fokas--Gel'fand formula for immersion}

\author{S Bertrand$^1$ and A M Grundland$^{2,3}$}

\address{$^1$ Department of Mathematics and Statistics, Universit\'e de Montr\'eal,\\ Montr\'eal CP 6128 Succ. Centre-Ville (QC) H3C 3J7, Canada}
\address{$^2$ Centre de Recherches Math\'ematiques, Universit\'e de Montr\'eal,\\ Montr\'eal CP 6128 Succ. Centre-Ville (QC) H3C 3J7, Canada}
\address{$^3$ Department of Mathematics and Computer Science, Universit\'e du Qu\'ebec, Trois-Rivi\`eres, CP 500 (QC) G9A 5H7, Canada}
\ead{bertrans@crm.umontreal.ca and grundlan@crm.umontreal.ca}
\vspace{10pt}

\begin{abstract}
In this paper, we construct and investigate two supersymmetric versions of the Fokas--Gel'fand formula for the immersion of 2D surfaces associated with a supersymmetric integrable system. The first version involves an infinitesimal deformation of the zero-curvature condition and the linear spectral problem associated with this system. This deformation leads the surfaces to be represented in terms of a bosonic supermatrix immersed in a Lie superalgebra. The second supersymmetric version is obtained by using a fermionic parameter deformation to construct surfaces expressed in terms of a fermionic supermatrix immersed in a Lie superalgebra. For both extensions, we provide a geometrical characterization of deformed surfaces using the super Killing form as an inner product and a super moving frame formalism. The theoretical results are applied to the supersymmetric sine-Gordon equation in order to construct super soliton surfaces associated with five different symmetries. We find integrated forms of these surfaces which represent constant Gaussian curvature surfaces and nonlinear Weingarten-type surfaces.
\end{abstract}

\pacs{12.60Jv, 02.20.Sv, 02.40.Ky}
\ams{35Q51, 53A05, 22E70}

\vspace{2pc}
\noindent{\it Keywords}: soliton surfaces, supersymmetric models, integrable systems.


\maketitle

\section{Introduction}\label{SecIntro}
In recent decades, an increasing number of supersymmetric (SUSY) extensions for quantum and classical models have been investigated (see e.g. \cite{BJ01,Crombrugghe,Henkel06,JP00}). In particular, super soliton solutions have been determined for SUSY extensions of various integrable systems of partial differential equations (PDEs), such as the SUSY sine-Gordon equation \cite{Aratyn,Coleman,Witten2,GHS09,Chaichian,Grammaticos,Gomes}, the SUSY Korteweg--de Vries equation \cite{Mathieu,MathieuLabelle,Liu,Chaichian}, the SUSY Schr\"odinger equations \cite{CIN02,Crombrugghe,Henkel06}, the SUSY Sawada--Kotera equations \cite{Tian} and the SUSY Hirota--Satsuma equations \cite{Popowicz97,DP04}. Super soliton solutions were obtained using  the connection between the super-Darboux transformations and the super-B\"acklund transformations (see e.g. \cite{Grammaticos,Matveev,Chaichian,Liu,Tian,Aratyn,Gomes,Siddiq06,GHS09} and references therein).

Supersymmetric versions of the equations of conformally parametrized surfaces provide rich classes of geometric objects \cite{Delisle16,DHZ15,BGH151,BGH152,Bertrand16}. In fact, until very recently, the formulation of two distinct SUSY extensions of the Gauss--Weingarten and Gauss--Codazzi (GC) equations for conformally parametrized surfaces immersed in a Grassmann superspace, one in terms of a bosonic superfield and the other in terms of a fermionic superfield, were the only known examples \cite{DHYZ,HZ06}.

On the other hand, the subalgebras of Lie point symmetries of the bosonic and fermionic SUSY GC equations were recently established in \cite{BGH151,BGH152}. The classification of the 1D subalgebras of each superalgebra into conjugacy classes has been performed. The symmetry reduction method was used to find invariants and reduced systems associated with the SUSY GC extensions \cite{BGH151,BGH152}. This approach allowed us to construct explicit solutions of these reduced SUSY systems, which correspond to different classes of surfaces immersed in a Grassmann superspace. These extensive results make conformally parametrized surfaces a rather special and interesting object of study. 

This paper is concerned with the investigation of different geometric aspects of these surfaces obtained in connection with integrable systems. Our main objective is to provide SUSY versions of immersion formulas for constructing large families of surfaces in Lie superalgebras linked with integrable SUSY GC equations. In order to achieve this goal we investigate certain features of point and generalized symmetries of SUSY integrable systems. The construction of the SUSY versions of the Fokas-Gel'fand (FG) formula for the immersion of 2D surfaces in Lie superalgebras is presented in detail. We demonstrate that a SUSY generalization of the classical main result on the immersion of 2D surfaces in a Lie algebra can be constructed. We show for these SUSY extensions that if there exists a common symmetry of the zero-curvature representation (ZCR) of an integrable system and its linear spectral problem (LSP) then the FG immersion formula is applicable in its original form. For this purpose, we write the SUSY version formula for immersion functions of 2D surfaces in Lie superalgebras in terms of vector fields and their prolongations rather than the notion of Fr\'echet derivatives \cite{FGFL,GP11}. In the classical case, the possibility of using a ZCR and its LSP to represent a moving frame on integrable surfaces has yielded many new results concerning the intrinsic geometric properties of such surfaces (see e.g. \cite{FGFL,Finkel,FinkelF}). The results obtained for the classical case were so promising that it seemed to be worthwhile to try to extend this method and check its effectiveness for the SUSY case. 

One of the purposes of this paper is to formulate a SUSY extension of the FG formula, which is obtained by applying a bosonic infinitesimal deformation to the potential matrices and the wavefunction of the LSP associated with the initial system in such a way that the deformed surface takes the form of a bosonic supermatrix. Next, another SUSY extension is derived using a fermionic parameter deformation of the potential matrices and the wavefunction, which implies that deformed surfaces take the form of a fermionic supermatrix. For both extensions, we provide geometrical characterizations of the deformed surfaces using the super Killing form as an inner product together with a SUSY version of the moving frame on the surface. These theoretical considerations are applied to the SUSY sine-Gordon equation. Surfaces associated with five different symmetries are investigated using the Sym--Tafel immersion formula, two gauge transformations and two Lie point symmetries. For each surface, we provide a geometric characterization via the Gaussian and mean curvatures based on the SUSY versions of the first and second fundamental forms. This is, in short, the aim of the paper.

The paper is organized as follows. In section 2, a brief exposition of the classical FG formula for the immersion of 2D surfaces in Lie algebras is presented for integrable systems. Section 3 contains an overview of the Grassmann algebra formalism and introduces the notation used in this paper. Section 4 is devoted to the construction of two SUSY versions of the FG formula for the immersion of 2D surfaces in Lie superalgebras. More specifically, in  section 4.1, we formulate the bosonic immersion of super soliton surfaces, while section 4.2 describes the fermionic immersion of super soliton surfaces. In section 5, surfaces associated with five different symmetries of the SUSY sine-Gordon equation are constructed, namely
the Sym--Tafel immersion formula, a bosonic gauge transformation, a bosonic symmetry deformation, a fermionic gauge transformation and a fermionic symmetry deformation, respectively. The conclusions and some possible future developments are presented in section 6.

\section{Immersion formula for soliton surfaces}\label{SecCla}
Consider an integrable system of PDEs 
\begin{equation}
\Delta[u]=0,\label{claPDE}
\end{equation}
in two independent variables $x_1,x_2$ and the dependent variables $u^k(x_1,x_2)$, which can be linearized by a matrix LSP given by
\begin{equation}
D_{x_\alpha}\Phi([u],\lambda)=U_\alpha([u],\lambda)\Phi([u],\lambda),\qquad \alpha=1,2.\label{claLSP}
\end{equation}
We use the abbreviated notation $[u]=(x_1,x_2,u^k,u^k_J)$ of an element of the jet space, where
\begin{equation*}
u^k_J=\frac{\partial^nu^k}{\partial x_{j_1}...\partial x_{j_n}},\qquad J=(j_1,...,j_n),\qquad \vert J\vert=n,\qquad j_i=1,2
\end{equation*}
with the total derivatives
\begin{equation*}
D_\alpha=\partial_{x_\alpha}+\sum_Ju^k_{J,\alpha}\frac{\partial}{\partial u^k_J},\qquad \alpha=1,2.
\end{equation*}
The compatibility conditions of (\ref{claLSP}) are in the form of a zero-curvature condition (ZCC) which is assumed to be valid for all values of the spectral parameter $\lambda\in\mathbb{C}$. This requirement implies that
\begin{equation*}
D_2U_1-D_1U_2+[U_1,U_2]=0,\label{claZCC}
\end{equation*}
which is equivalent to the original PDEs (\ref{claPDE}). It was shown \cite{Sym82,Sym85,Tafel95} that if a solution $\Phi([u],\lambda)$ of the LSP (\ref{claLSP}) is an element of a Lie group $G$ and $U_\alpha([u],\lambda)$ are functions in the associated Lie algebra $\mathfrak{g}$, then the function
\begin{equation}
F([u],\lambda)=\Phi^{-1}([u],\lambda)(\partial_\lambda\Phi([u],\lambda)),\label{claFst}
\end{equation}
where $\partial_\lambda$ is the partial derivative with respect to the spectral parameter $\lambda$, takes values in the Lie algebra $\mathfrak{g}$. The function $F$ can be interpreted for a fixed value of $\lambda$ as a surface in a Lie algebra $\mathfrak{g}$ provided that the tangent vectors
\begin{equation*}
D_\alpha F=\Phi^{-1}(\partial_\lambda U_\alpha)\Phi,\qquad\alpha=1,2
\end{equation*}
are linearly independent. Such a formula, which was first proposed by Sym \cite{Sym07} and Tafel \cite{Tafel95}, and subsequently used by many authors (see e.g. \cite{RS02,Bob}), allowed the link between classical geometry and integrable systems to be established, leading to the requirement that all 2D soliton solutions be represented by a one-parameter family of surfaces parametrized by the spectral parameter \cite{Bob}. Since then, the applicability of the Sym--Tafel formula for immersion to geometric problems of 2D surfaces related to integrable equations has been extended. In particular, new terms have been added to its original form (\ref{claFst}). As proven in \cite{FG96}, for any $\mathfrak{g}$-valued matrix functions $A_\alpha([u],\lambda)$, $\alpha=1,2$, which satisfy
\begin{equation}
D_2A_1-D_1A_2+[A_1,U_2]+[U_1,A_2]=0,\label{claIDZCC}
\end{equation}
there exists a $\mathfrak{g}$-valued immersion function $F$ with tangent vectors given by
\begin{equation*}
D_\alpha F=\Phi^{-1}A_\alpha\Phi,\qquad \alpha=1,2.
\end{equation*}
Whenever the matrix functions $A_1$ and $A_2$ are linearly independent, $F$ is an immersion function for a 2D surface in the Lie algebra $\mathfrak{g}$. As proven in \cite{GP11,FGFL}, three linearly independent terms which satisfy (\ref{claIDZCC}) are given by
\begin{equation*}
A_\alpha=\beta(\lambda)\partial_\lambda U_\alpha+(D_\alpha S+[S,U_\alpha])+\mbox{pr}\omega_RU_\alpha\in\mathfrak{g},\qquad\alpha=1,2
\end{equation*}
where $\beta(\lambda)$ is an arbitrary scalar function of $\lambda$, $S([u],\lambda)$ is an arbitrary $\mathfrak{g}$-valued function of $[u]$ and $\lambda$, and
\begin{equation*}
\omega_R=R^k[u]\partial_{u^k}
\end{equation*}
is the vector field, written in evolutionary form, of the generalized symmetry of the integrable PDEs (\ref{claPDE}), while 
\begin{equation*}
\mbox{pr}\omega_R=\omega_R+D_JR^k\partial_{u_J^k}
\end{equation*}
is the prolongation of vector field $\omega_R$. Furthermore, it has been proven in \cite{GP11,FGFL} that the $\mathfrak{g}$-valued function $F$ can be explicitly integrated as
\begin{equation}
F=\beta(\lambda)\Phi^{-1}(\partial_\lambda\Phi)+\Phi^{-1}S\Phi+\Phi^{-1}(\mbox{pr}\omega_R\Phi)\label{claF}
\end{equation}
as long as $\omega_R$ is a generalized symmetry of the integrable PDEs (\ref{claPDE}) and its LSP (\ref{claLSP}). The three terms in (\ref{claF}) correspond to conformal transformations of the spectral parameter $\lambda$ (the Sym--Tafel formula for immersion \cite{Sym82,Sym85,Tafel95}), a gauge symmetry of the LSP (due to Cieslinski and Doliwa \cite{Cieslinski07,DS92}) and a generalized common symmetry of the ZCC (\ref{claZCC}) and the LSP (\ref{claLSP}) (proposed by Fokas and Gel'fand \cite{FG96} and further developed in \cite{FGFL,GP11}).

The second term in (\ref{claF}), associated with the gauge symmetry of the LSP (\ref{claLSP}), can be integrated explicitly as
\begin{equation*}
F^S=\Phi^{-1}S([u],\lambda)\Phi\in\mathfrak{g},
\end{equation*}
which is consistent with the tangent vectors
\begin{equation*}
D_\alpha F^S=\Phi^{-1}(D_\alpha S+[S,U_\alpha])\Phi.\qquad \alpha=1,2
\end{equation*}
For $F$ to be an immersion function of a 2D surface, we require that the tangent vectors be linearly independent. Note that any surface $P\in\mathfrak{g}$ can be expressed as $P=\Phi^{-1}S\Phi=F^S$ and hence $F^S$ represents a completely arbitrary surface immersed in a Lie algebra $\mathfrak{g}$. So we can interpret the surface $F^S$ as an arbitrary surface immersed in the Lie algebra $\mathfrak{g}$ written in the frame defined by conjugation by the wavefunction $\Phi$, an element of the Lie group $G$.

The third term in (\ref{claF}) corresponds to the FG formula for immersion, which is applicable in its original form under the condition that the vector field $\Omega_R$ is a common symmetry of both the original system (\ref{claPDE}) and its LSP (\ref{claLSP}) \cite{Grundland16}. In this case the matrices
\begin{equation}
A_\alpha=\mbox{pr}\omega_RU_\alpha\label{claAfg}
\end{equation}
identically satisfy the determining equation (\ref{claIDZCC}). In the derivation of (\ref{claAfg}) we have used the fact that the total derivatives $D_\alpha$ commute with the prolongation of a vector field $\omega_R$ written in evolutionary form \cite{Olver}, that is
\begin{equation*}
[D_\alpha,\mbox{pr}\omega_R]=0,\qquad \alpha=1,2.
\end{equation*}
Thus, there exists a $\mathfrak{g}$-valued immersion function $F$ with tangent vectors
\begin{equation*}
D_\alpha F=\Phi^{-1}(\mbox{pr}\omega_RU_\alpha)\Phi.
\end{equation*}
Further, the immersion function $F$ can be integrated as
\begin{equation*}
F=\Phi^{-1}(\mbox{pr}\omega_R\Phi)\in\mathfrak{g},\label{claFfg}
\end{equation*}
if and only if the vector field $\omega_R$ is also a generalized symmetry of the LSP (\ref{claLSP}).

In sections \ref{SecB} and \ref{SecF}, the three terms in the immersion formula (\ref{claF}) will be used to construct two SUSY versions of the FG formula for the immersion of 2D surfaces in Lie superalgebras.

\section{Preliminaries on Grassmann algebras}\label{SecPre}
In this section, we present an overview of the definitions and formalism used throughout this paper. A more detailed description of Grassmann algebra can be found in \cite{Cornwell,DeWitt,Freed,Kac,Varadarajan,Berezin,Binetruy,Dine,Terning,Weinberg} and the references therein.

The complex Grassmann algebra $\mathbb{G}$ (denoted $\mathbb{C}B_L$ in \cite{Cornwell}) is a commutative associative algebra generated by a set of odd elements $\xi_j$ together with the unit $1$, where
\begin{equation*}
\xi_j\xi_k+\xi_k\xi_j=0\qquad\mbox{and}\qquad 1\xi_j=\xi_j.
\end{equation*}
Therefore any odd generator (or any odd element) of $\mathbb{G}$ satisfies the property
\begin{equation*}
\xi_j\xi_j=0 \qquad(\mbox{no summation}).
\end{equation*}
An odd element of $\mathbb{G}$ is composed of a linear combinaison of odd products of generators (e.g. $\xi_1+\xi_1\xi_2\xi_3$), while an even element of $\mathbb{G}$ is composed of a linear combinaison of even products of generators (e.g. $1+\xi_1\xi_2$).
The number of generators for each case is not specified, but we consider that there is a sufficient number of them to make all considered formulas meaningful. 
The degree of a homogeneous element $a\in\mathbb{G}$ is defined to be
\begin{equation*}
\deg(a)=\left\lbrace\begin{array}{cl}
0 & \mbox{for an even element,} \\
1 & \mbox{for an odd element.}
\end{array}\right.
\end{equation*}
One can also define the concept of a $(p+q)\times(r+s)$ supermatrix
\begin{equation*}
M=\left(\begin{array}{cc}
A & B \\
C & D
\end{array}\right),
\end{equation*}
where the submatrices $A$, $B$, $C$ and $D$ are of dimensions $p\times r$, $q\times r$, $p\times s$ and $q\times s$, respectively. The matrix $M$ is said to be an even supermatrix (or an even element of the Lie superalgebra $\mathfrak{gl}(p\vert q,\mathbb{G})$ if $r=p$ and $q=s$) if the submatrices $A$ and $D$ take their values in the even elements of $\mathbb{G}$ and if $B$ and $C$ take their values in the odd elements of $\mathbb{G}$. Conversely, the matrix $M$ is said to be an odd supermatrix (or an odd element of the Lie superalgebra $\mathfrak{gl}(p\vert q,\mathbb{G})$ if $r=p$ and $q=s$) if the the submatrices $A$ and $D$ take their values in the odd elements of $\mathbb{G}$ and if $B$ and $C$ take their values in the even elements of $\mathbb{G}$. The degree of a supermatrix is defined similarly to the degree of an element of $\mathbb{G}$, which is
\begin{equation*}
\deg(M)=\left\lbrace\begin{array}{cl}
0 & \mbox{if $M$ is an even supermatrix,} \\
1 & \mbox{if $M$ is an odd supermatrix.}
\end{array}\right.
\end{equation*}
The set of square $(p+q)\times(p+q)$ supermatrices with complex entries forms the Lie superalgebra $\mathfrak{gl}(p\vert q,\mathbb{G})$ and any Lie superalgebra $\mathfrak{g}$ has to satisfy the Lie super bracket
\begin{equation*}
M_1M_2-(-1)^{\deg(M_1)\deg(M_2)}M_2M_1=M_3\in\mathfrak{g}
\end{equation*}
for any $M_1,M_2\in\mathfrak{g}$. The Lie super bracket will be denoted by the commutator and anticommutator,
\begin{equation*}
\hspace{-1cm}[M_1,M_2]=M_1M_2-M_2M_1\qquad\mbox{and}\qquad\lbrace M_1,M_2\rbrace=M_1M_2+M_2M_1,
\end{equation*}
respectively, depending on the degree of $M_1$ and $M_2$. The associated Lie supergroup $GL(p\vert q,\mathbb{G})$ is composed of all (even) supermatrices of dimension $(p+q)\times(p+q)$ that are invertible.

In this paper, we use the convention that partial derivatives involving odd variables satisfy the Leibniz rule
\begin{equation*}
\partial_{\theta^j}(hg)=(\partial_{\theta^j}h)g+(-1)^{\deg(h)}h\partial_{\theta^j}g,
\end{equation*}
and
\begin{equation*}
f_{\theta^2\theta^1}=\partial_{\theta^1}(\partial_{\theta^2}f)=-\partial_{\theta^2}(\partial_{\theta^1}f)=-f_{\theta^1\theta^2}.
\end{equation*}
The partial derivatives with respect to odd coordinates change the parity of an even function to an odd function and vice versa.

In this paper, we do not follow the implicit notation for the odd derivative of a supermatrix (or multiplication by an odd scalar) used in \cite{Cornwell}, e.g.
\begin{equation*}
\partial_{\theta^j}M=\left(\begin{array}{cc}
\partial_{\theta^j}A & \partial_{\theta^j}B \\
-\partial_{\theta^j}C & -\partial_{\theta^j}D
\end{array}\right).
\end{equation*}
Therefore, we introduce the matrix $E$ such that
\begin{equation*}
E\partial_{\theta^j}M=\left(\begin{array}{cc}
\partial_{\theta^j}A & \partial_{\theta^j}B \\
-\partial_{\theta^j}C & -\partial_{\theta^j}D
\end{array}\right),\qquad E=\left(\begin{array}{cc}
I_p & 0 \\
0 & -I_q
\end{array}\right),
\end{equation*}
where $I_p$ is the $p\times p$ identity matrix. One should note that in this paper the terms even and odd are equivalent to bosonic and fermionic, respectively.

By considering the super Killing form $B:\mathfrak{gl}(p\vert q,\mathbb{G})\times\mathfrak{gl}(p\vert q,\mathbb{G})\rightarrow\mathbb{G}$ defined using the supertrace \cite{Cornwell},
\begin{equation}
\hspace{-2cm}\langle M,N\rangle=\alpha\mbox{ str}(MN)=\alpha\tr(E^{\deg(MN)+1}MN),\qquad E=\left(\begin{array}{cc}
I_p & 0 \\
0 & -I_q
\end{array}\right),\label{SKilling}
\end{equation}
where $\alpha$ is a nonzero real constant (e.g. in the examples of section \ref{SecEx}, $\alpha=1/2$) and $M,N$ are supermatrices in $\mathfrak{gl}(p\vert q,\mathbb{G})$, we can introduce an inner product $\langle\cdot,\cdot\rangle$, which has the following properties:
\begin{enumerate}
\item Left linearity,
\begin{equation*}
\langle M+N,P\rangle=\langle M,P\rangle+\langle N,P\rangle.
\end{equation*}
\item Right linearity,
\begin{equation*}
\langle M,N+P\rangle=\langle M,N\rangle+\langle M,P\rangle.
\end{equation*}
\item Inner permutation,
\begin{equation*}
\langle MN,P\rangle=\langle M,NP\rangle.
\end{equation*}
\item Outer permutation,
\begin{equation*}
\langle M,N\rangle=(-1)^{\deg(M)\deg(N)}\langle N,M\rangle.
\end{equation*}
\item Invariance under group conjugation,
\begin{equation*}
\langle S^{-1}MS,S^{-1}NS\rangle=\langle M,N\rangle,
\end{equation*}
for $S\in GL(p\vert q,\mathbb{G})$.
\item \label{prop6} Supercommutator,
\begin{equation*}
\langle M,[N,P]\rangle=\langle [M,N], P\rangle\label{PropB6}
\end{equation*}
or
\begin{equation*}
\langle M,\lbrace N,P\rbrace\rangle=\langle \lbrace M,N\rbrace, P\rangle,
\end{equation*}
depending on the degree of $M$, $N$ and $P$ for $\deg(M)=\deg(N)=\deg(P)$.
\end{enumerate}
One should note from property (\ref{prop6}) that the commutator/anticommutator acts as the vector product for the purpose of obtaining an orthogonal vector, e.g. 
\begin{equation*}
\langle M,[M,N]\rangle=\langle [M,M],N\rangle=0.
\end{equation*}

\section{SUSY versions of the Fokas--Gel'fand formula}\label{SecSUSY}
Let $\Delta[u]=0$ be an integrable system of PDEs in terms of the bosonic independent variables $x_1$ and $x_2$, the fermionic independent variables $\theta^1$ and $\theta^2$, and the SUSY dependent variables $u^k$ and their derivatives. Also, let us assume that there exists an associated LSP of the form
\begin{equation}
\Omega(\lambda,[u])=D_j\Psi(\lambda,[u])-U_j(\lambda,[u])\Psi(\lambda,[u])=0,\qquad j=1,2\label{SLSP}
\end{equation}
where $\Psi(\lambda,[u])\in GL(p\vert q,\mathbb{G})$, $U_j(\lambda,[u])$ are fermionic supermatrices in $\mathfrak{gl}(p\vert q,\mathbb{G})$, $\lambda$ is a spectral parameter and the covariant derivatives
\begin{equation*}
D_j=\partial_{\theta^j}-i\theta^j\partial_{x_j},\qquad j=1,2
\end{equation*}
satisfy the properties
\begin{equation*}
\lbrace D_1,D_2\rbrace=0,\qquad D_j^2=-i\partial_{x_j}.
\end{equation*}
The compatibility conditions of the LSP (\ref{SLSP}) (i.e. the ZCC) are given by
\begin{equation}
\hspace{-1cm}D_1U_2+D_2U_1-\lbrace EU_1,EU_2\rbrace=0,\qquad\mbox{where}\qquad E=\left(\begin{array}{cc}
I_p & 0 \\
0 & -I_q
\end{array}\right),\label{SZCC}
\end{equation}
which, for any value of $\lambda$, is equivalent to the original system of PDEs $\Delta[u]=0$. One should note that the fermionic derivatives
\begin{equation}
J_k=\partial_{\theta^k}+i\theta^k\partial_{x_k},\qquad k=1,2\label{J}
\end{equation}
anticommute with the differential generators $D_j$ and generate the SUSY transformations
\begin{equation*}
x_k\rightarrow x_k+i\gamma\theta^k,\qquad \theta^k\rightarrow\theta^k+i\gamma,\qquad k=1,2
\end{equation*}
where $\gamma$ is a odd-valued parameter.

\subsection{SUSY version of the Fokas--Gel'fand formula for bosonic immersion}\label{SecB}
We now consider an infinitesimal transformation on the potential supermatrices $U_j$,
\begin{equation}
\begin{array}{l}
\tilde{U}_1=U_1+\epsilon A_1,\qquad\tilde{U}_2=U_2+\epsilon A_2,
\end{array}\label{BdefU}
\end{equation}
where the matrices $A_j(\lambda,[u])$ are fermionic supermatrices in $\mathfrak{gl}(p\vert q,\mathbb{G})$ and $\epsilon$ is an infinitesimal bosonic parameter such that $\epsilon^2$ is negligeable. We also consider the infinitesimal transformation on the wavefunction $\Psi$ given by
\begin{equation}
\tilde{\Psi}=\Psi(I+\epsilon F).\label{BdefPsi}
\end{equation}
Assuming that these infinitesimal transformations preserve the LSP,
\begin{equation}
D_j\tilde{\Psi}=\tilde{U}_j\tilde{\Psi},\qquad j=1,2\label{BIDLSP}
\end{equation}
we obtain a deformed surface $F(\lambda,[u])$ expressed in terms of bosonic supermatrices of the Lie superalgebra $\mathfrak{gl}(p\vert q,\mathbb{G})$ under the condition that the tangent vectors are linearly independent. One can determine that the tangent vectors $ED_j(F)$ are
\begin{equation}
ED_j(F)=\Psi^{-1}EA_j\Psi.\label{BSvtan}
\end{equation}
Moreover, the compatibility conditions of the tangent vectors (\ref{BSvtan}) are equivalent to the infinitesimal transformation of the ZCC (\ref{SZCC}), which is
\begin{equation}
D_1A_2+D_2A_1-\lbrace EA_1,EU_2\rbrace-\lbrace EA_2,EU_1\rbrace=0.\label{BIDZCC}
\end{equation}
Let us consider the deformed surface (an analogue of the classical case, see equation (\ref{claF}))
\begin{equation}
F=\Psi^{-1}\beta(\lambda)(\partial_\lambda\Psi)+\Psi^{-1}ES\Psi+\Psi^{-1}(\mbox{pr}\omega\Psi).\label{BF}
\end{equation}
We use the spectral symmetry generator $\beta(\lambda)\partial_\lambda$, where $\beta(\lambda)$ is an arbitrary function such that $\deg(\beta)=\deg(\lambda)$ (i.e. if $\lambda$ is bosonic, then $\beta(\lambda)$ is bosonic and if $\lambda$ is fermionic, then $\beta(\lambda)$ is fermionic). The gauge $S(\lambda,[u])$ is an even supermatrix in $\mathfrak{gl}(p\vert q,\mathbb{G})$ and the bosonic generator $\omega$ spans a symmetry transformation for both the ZCC $\Delta[u]=0$ and the LSP $\Omega(\lambda,[u])=0$. Then, the supermatrices $A_j$ take the form
\begin{equation}
\hspace{-2cm}A_j=\beta(\lambda)\partial_\lambda U_j+E\left(D_jS+[ES,EU_j]\right)+\left(\mbox{pr}\omega U_j+\left([D_j,\mbox{pr}\omega]\Psi\right)\Psi^{-1}\right),\label{BA}
\end{equation}
which satisfy equation (\ref{BIDZCC}).

\begin{proposition}
Let us assume that there exists an LSP of the form (\ref{SLSP}) associated with a SUSY integrable system of PDEs $\Delta[u]=0$ such that the ZCC (\ref{SZCC}) is equivalent to $\Delta[u]=0$. If we consider the bosonic infinitesimal deformations (\ref{BdefU}) and (\ref{BdefPsi}) that preserve both the LSP and the ZCC, i.e. $A_1,A_2$ and $F$ must satisfy equations (\ref{BIDLSP})-(\ref{BIDZCC}), then there exists an immersion bosonic supermatrix $F$ which defines a 2D surface provided that its tangent vectors (\ref{BSvtan}) are linearly independent.
\end{proposition}

\begin{corollary}
If one considers a deformed surface $F$, as defined in proposition 1, of the form (\ref{BF}), then the linearly independent supermatrices $A_1$ and $A_2$ appearing in the tangent vectors (\ref{BSvtan}) take the form (\ref{BA}).
\end{corollary}

A geometric characterization of the deformed surface $F$ can lead us to a better understanding of the PDEs under investigation. However, an explicit solution for the wavefunction $\Psi$ can, in some cases, be a task harder to accomplish than to solve the initial PDEs. Therefore, by choosing an inner product which is invariant under the automorphism $\mathfrak{g}\rightarrow\Psi^{-1}\mathfrak{g}\Psi$, we can eliminate the wavefunction $\Psi$ and obtain a pseudo-Riemannian immersion formula. Throughout this paper, we consider the super Killing form defined in equation (\ref{SKilling}).

Using the super Killing form, we can define the coefficients of the first fundamental form to be
\begin{equation*}
g_{ij}=\langle ED_i(F),ED_j(F)\rangle=\langle EA_i,EA_j\rangle,
\end{equation*}
which are bosonic quantities. However, this inner product requires that the coefficients $g_{ii}$ be zero. In order to lift the degeneracy of this inner product, it is convenient to use the alternative definition
\begin{equation}
g_{ii}=\langle EA_i,A_j\rangle,\qquad
g_{12}=-g_{21}=\langle EA_1,EA_2\rangle,\label{BGg}
\end{equation}
which we use throughout the rest of the paper for the bosonic immersion.
For both definitions, the first fundamental form is given by
\begin{equation*}
I=(d_1)^2g_{11}+2d_1d_2g_{12}+(d_2)^2g_{22},\label{BSI}
\end{equation*}
where the $d_j$ are fermionic differential forms which are the infinitesimal displacement in the direction of $D_j$ and satisfy the relation
\begin{equation*}
\lbrace d_1,d_2\rbrace=0.
\end{equation*}
These operators are defined as \cite{BGH151}
\begin{equation}
d_j=d\theta^j+idx_j\partial_{\theta^j}.\qquad j=1,2\label{SUSYdj}
\end{equation}
In order to construct the second fundamental form, we introduce a unit normal vector $N$ in terms of a bosonic supermatrix which has the properties
\begin{equation*}
\langle N,N\rangle=1,\qquad\langle ED_jF,N\rangle=0,\qquad j=1,2.
\end{equation*}
A unit normal vector can be given by
\begin{equation*}
N=\frac{\lbrace ED_1(F),ED_2(F)\rbrace}{\langle\lbrace ED_1(F),ED_2(F)\rbrace,\lbrace ED_1(F),ED_2(F)\rbrace\rangle^{1/2}},
\end{equation*}
or equivalently
\begin{equation}
N=\frac{\Psi^{-1}\lbrace EA_1,EA_2\rbrace\Psi}{\langle\lbrace EA_1,EA_2\rbrace,\lbrace EA_1,EA_2\rbrace\rangle^{1/2}}.\label{BSN}
\end{equation}
Therefore, the coefficients of the second fundamental form are given by the bosonic quantities
\begin{equation}
b_{ij}=\langle D_jD_iF,N\rangle=\langle D_jA_i-\lbrace EA_i,EU_j\rbrace,\Psi N\Psi^{-1}\rangle,\label{Bbij}
\end{equation}
which have the property $b_{12}=-b_{21}$. The second fundamental form is
\begin{equation*}
I\hspace{-0.1cm}I=(d_1)^2b_{11}+2d_1d_2b_{12}+(d_2)^2b_{22}.\label{BSII}
\end{equation*}
The Gaussian and mean curvatures are given, respectively, by
\begin{equation}
K=\frac{b_{11}b_{22}-b_{12}b_{21}}{g_{11}g_{22}-g_{12}g_{21}}=\frac{b_{11}b_{22}+(b_{12})^2}{g_{11}g_{22}+(g_{12})^2},\label{BSK}
\end{equation}
\begin{equation}
H=\frac{b_{11}g_{22}+b_{22}g_{11}-b_{12}g_{21}-b_{21}g_{12}}{2(g_{11}g_{22}-g_{12}g_{21})}=\frac{b_{11}g_{22}+b_{22}g_{11}+2b_{12}g_{12}}{2(g_{11}g_{22}+(g_{12})^2)},\label{BSH}
\end{equation}
where both curvatures are bosonic quantities. One should note that the coefficients $g_{ij}$ and $b_{ij}$ are explicitly given using the supermatrices $U_j$, $A_j$ and $N$.

\subsection{SUSY version of the Fokas--Gel'fand formula for fermionic immersion}\label{SecF}
We now consider a transformation on the potential supermatrices $U_j$,
\begin{equation}
\tilde{U}_1=U_1+\epsilon EA_1,\qquad\tilde{U}_2=U_2+\epsilon EA_2,\label{FdefU}
\end{equation}
where $\epsilon$ is a fermionic parameter, the matrices $A_j(\lambda,[u])$ are bosonic supermatrices in $\mathfrak{gl}(p\vert q,\mathbb{G})$, together with the transformation on the wavefunction $\Psi$,
\begin{equation}
\tilde{\Psi}=\Psi(I+\epsilon EF),\label{FdefPsi}
\end{equation}
such that the LSP remains invariant under these transformations, i.e.
\begin{equation}
D_j\tilde{\Psi}=\tilde{U}_j\tilde{\Psi},\qquad j=1,2.\label{FIDLSP}
\end{equation}
We obtain a deformed surface $F(\lambda,[u])$ expressed in terms of fermionic supermatrices in the Lie superalgebra $\mathfrak{gl}(p\vert q,\mathbb{G})$ assuming that the tangent vectors are linearly independent. The tangent vectors $ED_j(F)$ are given by
\begin{equation}
ED_j(F)=-\Psi^{-1}EA_j\Psi,\label{FSvtan}
\end{equation}
up to the addition of a bosonic supermatrix $R$ such that $\epsilon R=0$.
The compatibility conditions of the tangent vectors (\ref{FSvtan}) are equivalent to the deformation of the ZCC (\ref{SZCC}), which is given by
\begin{equation}
D_1A_2+D_2A_1+[EA_1,EU_2]+[EA_2,EU_1]=0.\label{FIDZCC}
\end{equation}
If one considers the deformed surface, one obtains
\begin{equation}
F=\Psi^{-1}E\beta(\lambda)(\partial_\lambda\Psi)+\Psi^{-1}ES\Psi+\Psi^{-1}E(\mbox{pr}\omega\Psi).\label{FF}
\end{equation}
We use the spectral symmetry generator $\beta(\lambda)\partial_\lambda$, where $\beta(\lambda)$ is an arbitrary function such that $\deg(\beta)=\deg(\lambda)+1$ mod $2$ (i.e. if $\lambda$ is bosonic, then $\beta(\lambda)$ is fermionic and if $\lambda$ is fermionic, then $\beta(\lambda)$ is bosonic). The gauge $S(\lambda,[u])$ is a fermionic supermatrix in $\mathfrak{gl}(p\vert q,\mathbb{G})$ and the fermionic generator $\omega$ spans a symmetry transformation for both the ZCC $\Delta[u]=0$ and the LSP $\Omega(\lambda,[u])=0$. Then, the supermatrices $A_j$ take the form
\begin{equation}
\hspace{-2cm}A_j=E\beta(\lambda)\partial_\lambda U_j-E\left(D_jS-\lbrace ES,EU_j\rbrace\right)+E\left(\mbox{pr}\omega U_j-\left(\lbrace D_j,\mbox{pr}\omega\rbrace\Psi\right)\Psi^{-1}\right),\label{FA}
\end{equation}
which satisfy equation (\ref{FIDZCC}).

\begin{proposition}
Let us assume that there exists an LSP of the form (\ref{SLSP}) associated with a SUSY integrable system of PDEs $\Delta[u]=0$ such that the ZCC (\ref{SZCC}) is equivalent to $\Delta[u]=0$. If we consider the fermionic parameter deformations (\ref{FdefU}) and (\ref{FdefPsi}) that preserve both the LSP and the ZCC, i.e. $A_1,A_2$ and $F$ must satisfy equations (\ref{FIDLSP})-(\ref{FIDZCC}), then there exists an immersion fermionic supermatrix $F$ which defines a 2D surface provided that its tangent vectors (\ref{FSvtan}) are linearly independent.
\end{proposition}

\begin{corollary}
If one considers a deformed surface $F$, as defined in proposition 2, of the form (\ref{FF}), then the linearly independent supermatrices $A_1$ and $A_2$ appearing in the tangent vectors (\ref{FSvtan}) take the form (\ref{FA}).
\end{corollary}

Using the inner product (\ref{SKilling}), we define the coefficients of the first fundamental form to be the bosonic quantities
\begin{equation}
g_{ij}=\langle ED_i(F),ED_j(F)\rangle=\langle EA_i,EA_j\rangle,\label{Fgij}
\end{equation}
with the property $g_{12}=g_{21}$. The first fundamental form is given by
\begin{equation*}
I=(d_1)^2g_{11}+2d_1d_2g_{12}+(d_2)^2g_{22},\label{FSI}
\end{equation*}
where the $d_j$ are fermionic differential forms which represent the infinitesimal displacement in the direction of $D_j$ and satisfy the relation
\begin{equation*}
\lbrace d_1,d_2\rbrace=0.
\end{equation*}
These operators are defined as in (\ref{SUSYdj}). In order to construct the second fundamental form, we must find a unit normal vector $N$ in terms of a bosonic supermatrix which has the properties
\begin{equation*}
\langle N,N\rangle=1,\qquad\langle ED_jF,N\rangle=0,\qquad j=1,2.
\end{equation*}
A unit normal vector is
\begin{equation*}
N=\frac{[ED_1(F),ED_2(F)]}{\langle [ED_1(F),ED_2(F)],[ED_1(F),ED_2(F)]\rangle^{1/2}},
\end{equation*}
or equivalently
\begin{equation}
N=\frac{\Psi^{-1}[ EA_1,EA_2]\Psi}{\langle [EA_1,EA_2],[EA_1,EA_2]\rangle^{1/2}},\label{FSN}
\end{equation}
whenever the division is possible.
Therefore, the coefficients of the second fundamental form are given by
\begin{equation}
b_{ij}=\langle D_jD_iF,N\rangle=\left\langle -(D_jA_i+ [EA_i,EU_j]),\Psi N\Psi^{-1}\right\rangle,\label{Fbij}
\end{equation}
which have the property $b_{12}=-b_{21}$ and are fermionic quantities. The second fundamental form is
\begin{equation*}
I\hspace{-1mm}I=(d_1)^2b_{11}+2d_1d_2b_{12}+(d_2)^2b_{22}.\label{FSII}
\end{equation*}
The Gaussian and mean curvatures are given, respectively, by
\begin{equation}
K=\frac{b_{11}b_{22}-b_{12}b_{21}}{g_{11}g_{22}-g_{12}g_{21}}=\frac{b_{11}b_{22}+(b_{12})^2}{g_{11}g_{22}-(g_{12})^2},\label{FSK}
\end{equation}
\begin{equation}
H=\frac{b_{11}g_{22}+b_{22}g_{11}-b_{12}g_{21}-b_{21}g_{12}}{2(g_{11}g_{22}-g_{12}g_{21})}=\frac{b_{11}g_{22}+b_{22}g_{11}}{2(g_{11}g_{22}-(g_{12})^2)},
\label{FSH}
\end{equation}
where $K$ is a bosonic quantity and $H$ is a fermionic one and such that they can be computed only using the supermatrices $U_j$, $A_j$ and $N$.

\section{Example: The SUSY sine-Gordon equation}\label{SecEx}
In this section, we apply the theory described in the previous sections to the SUSY sine-Gordon equation. The SUSY sine-Gordon equation takes the form \cite{Chaichian,Grammaticos,Gomes}
\begin{equation}
D_2D_1\phi=i\sin\phi,\label{sG}
\end{equation}
where $\phi$ is a bosonic superfunction of $x_1$, $x_2$, $\theta^1$ and $\theta^2$, which can be decomposed as the truncated series
\begin{equation*}
\phi=\phi_0(x_1,x_2)+\phi_1(x_1,x_2)\theta^1+\phi_2(x_1,x_2)\theta^2+\phi_{12}(x_1,x_2)\theta^1\theta^2,
\end{equation*}
where $\phi_0$ and $\phi_{12}$ are bosonic functions of $x_1,x_2$ and $\phi_1$ and $\phi_2$ are fermionic functions of $x_1,x_2$. The associated LSP \cite{Siddiq06,BGH152}
\begin{equation}
\hspace{-2.5cm}\begin{array}{c}
D_j\Psi=U_j\Psi,\qquad j=1,2\\
U_1=\frac{1}{2\sqrt{\lambda}}\left(\begin{array}{ccc}
0 & 0 & ie^{i\phi} \\
0 & 0 & -ie^{-i\phi} \\
-e^{-i\phi} & e^{i\phi} & 0
\end{array}\right),\qquad U_2=\left(\begin{array}{ccc}
iD_2\phi & 0 & -i\sqrt{\lambda} \\
0 & -iD_2\phi &i\sqrt{\lambda} \\
-\sqrt{\lambda} & \sqrt{\lambda} & 0
\end{array}\right),
\end{array}\label{sGU}
\end{equation}
with the bosonic spectral parameter $\lambda$ for which the ZCC satisfies
\begin{equation*}
D_1U_2+D_2U_1-\lbrace EU_1,EU_2\rbrace=0,\qquad E=\left(\begin{array}{ccc}
1 & 0 & 0 \\
0 & 1 & 0 \\
0 & 0 & -1
\end{array}\right)\label{sGZCC}
\end{equation*}
is equivalent to the SUSY sine-Gordon equation (\ref{sG}) for any value of $\lambda$. 

One way to indroduce the spectral parameter in the potential supermatrices $U_1$ and $U_2$ is to consider the linear problem
\begin{equation}
D_j\hat{\Psi}=\hat{U}_j\hat{\Psi}, \qquad j=1,2\label{sGLP}
\end{equation}
where the matrices $\hat{U}_j$ take the form
\begin{equation*}
\hspace{-1cm}\hat{U}_1=\frac{1}{2}\left(\begin{array}{ccc}
0 & 0 & ie^{i\phi} \\
0 & 0 & -ie^{-i\phi} \\
-e^{-i\phi} & e^{i\phi} & 0
\end{array}\right),\qquad \hat{U}_2=\left(\begin{array}{ccc}
iD_2\phi & 0 & -i \\
0 & -iD_2\phi &i \\
-1 & 1 & 0
\end{array}\right).
\end{equation*}
From there, one should note that the vector field
\begin{equation}
\omega=2x_1\partial_{x_1}-2x_2\partial_{x_2}+\theta^1\partial_{\theta^1}-\theta^2\partial_{\theta^2},\label{sGvec}
\end{equation}
is a symmetry generator of the SUSY sine-Gordon equation, but not of the linear problem (\ref{sGLP}). This vector field (\ref{sGvec}) generates the transformations
\begin{equation*}
\hspace{-2.5cm}\tilde{x}_+=\lambda x_+,\qquad\tilde{x}_-=\lambda^{-1}x_-,\qquad\tilde{\theta}^+=\lambda^{1/2}\theta^+,\qquad\tilde{\theta}^-=\lambda^{-1/2}\theta^-,\qquad \lambda=\pm e^{\mu},\label{transformation}
\end{equation*}
where $\mu$ is a bosonic-valued parameter. By imposing these transformations on the supermatrices $\hat{U}_1$ and $\hat{U}_2$ we obtain, after some computation, the potential supermatrices defined in (\ref{sGU}). Therefore, the parameter $\lambda$ can play the role of the spectral parameter \cite{BGH152}.

In the following three examples we apply the theory from section \ref{SecB} to the SUSY sine-Gordon equation and we derive the bosonic immersion of surfaces. In the remaining two examples, we consider the fermionic immersion described in section \ref{SecF}.

\subsection{Sym--Tafel formula for a bosonic immersion}
In this case, we consider the deformed surface that takes the form of the bosonic supermatrix
\begin{equation}
F=\Psi^{-1}\beta(\lambda)\partial_\lambda\Psi\in\mathfrak{sl}(2\vert1,\mathbb{G}),\label{FST}
\end{equation}
with tangent vectors given by
\begin{equation*}
ED_jF=\Psi^{-1}E\beta(\lambda)\partial_\lambda U_j\Psi=\Psi^{-1}EA_j\Psi
\end{equation*}
for a bosonic arbitrary function $\beta(\lambda)$. Explicitly, the matrices $A_j$ are linearly independent and take the form
\begin{equation*}
\hspace{-2.5cm}A_1=\frac{-\beta}{4\sqrt{\lambda^3}}\left(\begin{array}{ccc}
0 & 0 & ie^{i\phi} \\
0 & 0 & -ie^{-i\phi} \\
-e^{-i\phi} & e^{i\phi} & 0
\end{array}\right)=\frac{-\beta}{2\lambda}U_1,\qquad A_2=\frac{\beta}{2\sqrt{\lambda}}\left(\begin{array}{ccc}
0 & 0 & -i \\
0 & 0 & i \\
-1 & 1 & 0
\end{array}\right)
\end{equation*}
and from equations (\ref{BGg}) we determine the coefficients of the first fundamental form
\begin{equation*}
\begin{array}{l}
g_{11}=\langle EA_1,A_1\rangle=\frac{-i\beta^2}{8\lambda^3},\\
g_{12}=-g_{21}=\langle EA_1,EA_2\rangle=-\frac{i\beta^2}{4\lambda^2}\cos\phi,\\
g_{22}=\langle EA_2,A_2\rangle=\frac{i\beta^2}{2\lambda},
\end{array}
\end{equation*}
such that 
\begin{equation*}
g=g_{11}g_{22}-g_{12}g_{21}=\frac{\beta^2}{16\lambda}\sin^2\phi.
\end{equation*}
In order to obtain the coefficients of the second fundamental form and the Gaussian and mean curvatures, we first need to compute a unit normal vector $N$ in matrix form, as given by equation (\ref{BSN}),
\begin{equation*}
N=\Psi^{-1}\left(\begin{array}{ccc}
-1 & 0 & 0 \\
0 & 1 & 0 \\
0 & 0 & 0
\end{array}\right)\Psi.
\end{equation*}
Therefore, we obtain that from equation (\ref{Bbij}) the coefficients of the second fundamental form are given by
\begin{equation*}
b_{11}=0,\qquad b_{12}=\frac{\beta}{2\lambda}\sin\phi,\qquad b_{22}=0,
\end{equation*}
and consequently the Gaussian and mean curvatures are given by
\begin{equation*}
K=\frac{4\lambda^2}{\beta^2},\qquad\mbox{and}\qquad H=\frac{-2i\lambda}{\beta}\cot\phi,
\end{equation*}
which we obtain from equations (\ref{BSK}) and (\ref{BSH}), respectively.
It should be noted that the Gaussian curvature is constant as for the classical case \cite{CFG00}, but the sign of the Gaussian curvature differs. By analogy with the classical geometry, if we look for umbilic points using the formula
\begin{equation*}
0=H^2-K=\frac{-4\lambda^2}{\beta^2}\csc^2\phi
\end{equation*}
we observe that umbilic points do not exist on this surface. Moreover, if we consider a SUSY version of the Euler-Poincar\'e character
\begin{equation*}
\chi=\frac{1}{2\pi}\int\int_\Omega d_2d_1g^{1/2}K,
\end{equation*}
and since soliton solutions of the SUSY sine-Gordon equation rapidly decay to zero, the Euler-Poincar\'e character vanishes,
\begin{equation*}
\chi=\frac{1}{2\pi}\int_{-\infty}^{\infty}d_2\int_{-\infty}^{\infty}d_1\sin\phi =\frac{-i}{2\pi}\int_{-\infty}^{\infty}d_2\int_{-\infty}^\infty d_1 (D_2D_1\phi)=0.
\end{equation*}
In analogy with the lemma 3.5 in \cite{CFG00} for soliton solutions of the classical sine-Gordon equation, this demonstrates that the SUSY version of this lemma is still valid. If the soliton solutions of the SUSY sine-Gordon equation (\ref{sG}) satisfy the conditions that the function $\phi$ and its derivatives tend to zero as the independent variables go to infinity, then we have
\begin{equation*}
\int_{-\infty}^\infty d_i\sqrt{g}K=0.\qquad i=1,2
\end{equation*}
In comparison with the classical geometry, if the deformed surface (\ref{FST}) is compact and connected, then it is homeomorphic to a torus since the Euler-Poincar\'e character vanishes \cite{DP97}.

\subsection{Bosonic gauge transformation}
Using the bosonic gauge supermatrix
\begin{equation*}
S=U_2D_2\phi=\sqrt{\lambda}D_2\phi\left(\begin{array}{ccc}
0 & 0 & -i \\
0 & 0 & i \\
-1 & 1 & 0
\end{array}\right)\in\mathfrak{sl}(2\vert1,\mathbb{G})
\end{equation*}
for the deformed surface
\begin{equation}
F=\Phi^{-1}ES\Psi\in\mathfrak{sl}(2\vert1,\mathbb{G})
\end{equation}
consistent with the linearly independent tangent vectors
\begin{equation*}
ED_jF=\Psi^{-1}EA_j\Psi,\\
\end{equation*}
where
\begin{equation*}
\begin{array}{l}
A_1=\left(\begin{array}{ccc}
-i\cos\phi D_2\phi & ie^{i\phi}D_2\phi & -\sqrt{\lambda}\sin\phi \\
ie^{-i\phi}D_2\phi & -i\cos\phi D_2\phi & \sqrt{\lambda}\sin\phi \\
-i\sqrt{\lambda}\sin\phi & i\sqrt{\lambda}\sin\phi & -2i\cos\phi D_2\phi
\end{array}\right),\\
\\
A_2=-i\sqrt{\lambda}\partial_{x_2}\phi\left(\begin{array}{ccc}
0 & 0 & -i \\
0 & 0 & i \\
1 & -1 & 0
\end{array}\right),
\end{array}
\end{equation*}
we get the metric coefficients
\begin{equation*}
g_{11}=2i\lambda\sin^2\phi,\qquad g_{12}=0,\qquad g_{22}=2i\lambda(\partial_{x_2}\phi)^2.
\end{equation*}
A unit normal vector $N$ is given by
\begin{equation*}
\hspace{-1cm}N=\Psi^{-1}\left(\begin{array}{ccc}
1 & 0 & 0 \\
0 & 1 & 0 \\
0 & 0 & -1
\end{array}\right)\Psi.
\end{equation*}
The coefficients of the second fundamental form reduce, after some straightfoward computation, to 
\begin{equation*}
\hspace{-1cm}b_{11}=2\sin\phi(iD_1\phi D_2\phi+\cos\phi),\qquad b_{12}=-2\cos\phi\partial_{x_2}\phi,\qquad b_{22}=0.
\end{equation*}
Therefore, we get the non-trivial Gaussian and mean curvatures
\begin{equation*}
K=\frac{-\cot^2\phi}{\lambda^2},\qquad\mbox{and}\qquad H=\frac{iD_2\phi D_1\phi-\cos\phi}{2\lambda\sin\phi}.
\end{equation*}
This surface represents a nonlinear Weingarten surface since there exists a second-order polynomial relation in $H$ with coefficients depending on $K$,
\begin{equation*}
f(K,H)=H^2-\frac{i}{2}\sqrt{K}H+2K=0.
\end{equation*}

\subsection{Fokas--Gel'fand formula for bosonic immersion}
If we consider the bosonic supermatrix
\begin{equation}
F=\Psi^{-1}(\mbox{pr}\omega\Psi)\in\mathfrak{sl}(2\vert1,\mathbb{G}),\label{F53}
\end{equation}
with a bosonic symmetry generator of both the SUSY sine-Gordon equation and its LSP given by $\partial_{x_1}$ (which generates a translation in the $x_1$ direction) \cite{BGH152}, then we obtain the tangent vectors given by
\begin{equation*}
ED_jF=\Psi^{-1}E\partial_{x_1}U_j\Psi=\Psi^{-1}EA_j\Psi,
\end{equation*}
where 
\begin{equation*}
\hspace{-2.5cm}
A_1=\frac{\partial_{x_1}\phi}{2\sqrt{\lambda}}\left(\begin{array}{ccc}
0 & 0 & -e^{i\phi} \\
0 & 0 & -e^{-i\phi} \\
i e^{-i\phi} & ie^{i\phi} & 0
\end{array}\right),\qquad
A_2=i\partial_{x_1}D_2\phi\left(\begin{array}{ccc}
1 & 0 & 0\\
0 & -1 & 0\\
0 & 0 & 0
\end{array}\right).
\end{equation*}
It is interesting to note that the metric coefficients degenerate to a curve-like metric, i.e.
\begin{equation*}
g_{11}=\frac{-i}{2\lambda}(\partial_{x_1}\phi)^2,\qquad g_{12}=0,\qquad g_{22}=-(\partial_{x_1}D_2\phi)^2=0.
\end{equation*}
A unit normal vector $N$ is given by
\begin{equation*}
N=\Psi^{-1}E\Psi,
\end{equation*}
which leads to the following coefficients of the second fundamental form
\begin{equation*}
b_{11}=b_{12}=b_{21}=0,\qquad b_{22}=2(\mbox{pr}\omega_RD_2\phi)D_2\phi.
\end{equation*}
It is also interesting to consider the isotropic normal vector $N$ of the form
\begin{equation*}
N=\Psi^{-1}\frac{-i(\partial_{x_1}\phi)^2}{2\lambda}\left(\begin{array}{ccc}
0 & 0 & -e^{i\phi} \\
0 & 0 & e^{-i\phi} \\
-ie^{-i\phi} & ie^{i\phi} & 0
\end{array}\right)\Psi,
\end{equation*}
so that we get a non-trivial second fundamental form
\begin{equation*}
II=(d_1)^2\frac{-\partial_{x_1}\phi D_1\phi}{\sqrt{2i\lambda}}+d_1d_2\frac{\sqrt{2}\partial_{x_1}D_2\phi}{\sqrt{i\lambda}}.
\end{equation*}
Consequently, we find the curvatures
\begin{equation*}
K=\frac{-1}{(\partial_{x_1}\phi)^2},\qquad\mbox{and}\qquad H=-\left(\frac{i\lambda}{2}\right)^{\frac{1}{2}}\frac{D_1\phi}{\partial_{x_1}\phi}.
\end{equation*}
Such an example of an isotropic normal vector has been obtained and investigated for the classical FG immersion formula \cite{GP12}. Note that for both cases the tangent vectors are linearly independent. So the immersion (\ref{F53}) defines a surface instead of a curve.

\subsection{Fermionic gauge transformation}
We consider the fermionic gauge supermatrix
\begin{equation*}
S=\left(\begin{array}{ccc}
-iD_2\phi & 0 & i\sqrt{\lambda} \\
0 & iD_2\phi & -i\sqrt{\lambda} \\
-\sqrt{\lambda} & \sqrt{\lambda} & 0
\end{array}\right)\in\mathfrak{sl}(2\vert1,\mathbb{G})
\end{equation*}
for the deformed surface
\begin{equation}
F=\Psi^{-1}ES\Psi\in\mathfrak{sl}(2\vert1,\mathbb{G}),
\end{equation}
which is consistent with the tangent vectors
\begin{equation*}
ED_jF=-\Psi^{-1}EA_j\Psi=\Psi^{-1}(D_jS-\lbrace ES,EU_j\rbrace)\Psi,
\end{equation*}
where the linearly independent matrices $A_j$ take the form
\begin{equation*}
\begin{array}{l}
A_1=\left(\begin{array}{ccc}
-ie^{-i\phi} & -ie^{i\phi} & e^{i\phi}D_2\phi/2\sqrt{\lambda} \\
-ie^{-i\phi} & -ie^{i\phi} & e^{-i\phi}D_2\phi/2\sqrt{\lambda} \\
ie^{-i\phi}D_2\phi/2\sqrt{\lambda} & ie^{i\phi}D_2\phi/2\sqrt{\lambda} & 2i\cos\phi
\end{array}\right),\\
~\\
A_2=\left(\begin{array}{ccc}
\partial_{x_2}\phi & 0 & -2\sqrt{\lambda}D_2\phi \\
0 & -\partial_{x_2}\phi & -2\sqrt{\lambda}D_2\phi \\
0 & 0 & 0
\end{array}\right),
\end{array}
\end{equation*}
such that the tangent vector $ED_1F$ is isotropic.
The first fundamental form coefficients are computed from equation (\ref{Fgij}) which give
\begin{equation*}
\begin{array}{c}
g_{11}=\langle EA_1,EA_1\rangle=0,\qquad  g_{22}=\langle EA_2,EA_2\rangle=(\partial_{x_2}\phi)^2,\\
g_{12}=\langle EA_1,EA_2\rangle=-i\sin\phi \partial_{x_2}\phi
\end{array}
\end{equation*}
and the unit normal vector $N$ from equation (\ref{FSN}) is given by
\begin{equation*}
\hspace{-2.5cm}N=\Psi^{-1}\left(\begin{array}{ccc}
0 & ie^{i\phi} & 2D_2\phi(-i\sqrt{\lambda}e^{-i\phi}/\partial_{x_2}\phi-e^{i\phi}/8\sqrt{\lambda}) \\
-ie^{-i\phi} & 0 & 2D_2\phi(-i\sqrt{\lambda}e^{i\phi}/\partial_{x_2}\phi+e^{-i\phi}/8\sqrt{\lambda}) \\
-ie^{-i\phi}D_2\phi/4\sqrt{\lambda} & ie^{i\phi}D_2\phi/4\sqrt{\lambda} & 0
\end{array}\right)\Psi.
\end{equation*}
Therefore, the coefficients of the second fundamental form are given by (\ref{Fbij})
\begin{equation*}
\hspace{-2.5cm}\begin{array}{c}
b_{11}=-iD_1\phi-\frac{i}{2\lambda}D_2\phi\cos\phi+\frac{iD_2\phi}{\partial_{x_2}\phi}\sin\phi+\frac{2iD_2\phi}{\partial_{x_2}\phi}\sin\phi\cos2\phi+3i\frac{D_2\phi}{\partial_{x_2}\phi}\sin2\phi\cos\phi,\\
b_{12}=-iD_2\phi(1+\cos2\phi+\frac{1}{2}\sin^2\phi),\qquad b_{22}=0,
\end{array}
\end{equation*}
which leads us to the Gaussian and mean curvatures of the fermionic immersion,
\begin{equation*}
K=0,\qquad H=\frac{b_{11}}{\sin^2\phi}.
\end{equation*}
This surface admits parabolic points and is a nonlinear Weingarten surface since $H$ is fermionic and such a relation holds
\begin{equation*}
f(K,H)=H^2+\alpha K=0,
\end{equation*}
where $\alpha$ is an arbitrary bosonic constant.

\subsection{Fokas--Gel'fand formula for fermionic immersion}
We investigate the deformed surface $F$ generated by the fermionic differential operator
\begin{equation*}
\omega_k=J_k=\partial_{\theta^k}+i\theta^k\partial_{x_k},\qquad k=1,2
\end{equation*}
where $J_k$ is defined in equation (\ref{J}) and is a symmetry of both the SUSY sine-Gordon equation and its LSP \cite{BGH152}.
The deformed surface
\begin{equation}
F=\Psi^{-1}E(\mbox{pr}\omega_k\Psi)\in\mathfrak{sl}(2\vert1,\mathbb{G})\label{F55}
\end{equation}
has tangent vectors of the form
\begin{equation*}
\begin{array}{l}
ED_jF=-\Psi^{-1}EA_j\Psi,\\
A_1=\frac{\mbox{pr}\omega_k\phi}{2\sqrt{\lambda}}\left(\begin{array}{ccc}
0 & 0 & -e^{i\phi} \\
0 & 0 & -e^{-i\phi} \\
-ie^{-i\phi} & -ie^{i\phi} & 0
\end{array}\right),\\
A_2=i\mbox{pr}\omega_kD_2\phi\left(\begin{array}{ccc}
1 & 0 & 0 \\
0 &-1 & 0 \\
0 & 0 & 0
\end{array}\right),
\end{array}
\end{equation*}
which are linearly independent.
One should note that the matrix $A_1$ is an isotropic vector. 
The metric coefficients are curve-like, i.e.
\begin{equation*}
g_{11}=0,\qquad g_{12}=0,\qquad g_{22}=-(\mbox{pr}\omega_k(D_2\phi))^2.
\end{equation*}
A unit normal vector $N$ is given by
\begin{equation*}
N=\Psi^{-1}\left(\begin{array}{ccc}
1 & 1 & 0 \\
-1 & 1 & 0 \\
0 & 0 & 1
\end{array}\right)\Psi
\end{equation*}
and the coefficients of the second fundamental form are
\begin{equation*}
b_{11}=\frac{\mbox{pr}\omega_R\phi}{2\lambda}\cos2\phi,\qquad b_{12}=b_{21}=b_{22}=0.
\end{equation*}
We can also use the isotropic normal vector $N$ given by
\begin{equation*}
N=\Psi^{-1}\frac{1}{\sqrt{2i}}\left(\begin{array}{ccc}
0 & 0 & e^{i\phi} \\
0 & 0 & -e^{-i\phi} \\
ie^{-i\phi} & -ie^{i\phi} & 0
\end{array}\right)\Psi,
\end{equation*}
and the coefficients of the second fundamental form do not degenerate to a curve-like form with the coefficients
\begin{equation*}
b_{11}=\frac{-D_1\phi(\mbox{pr}\omega_k\phi)}{\sqrt{2i\lambda}},\qquad b_{12}=\frac{-i\mbox{pr}\omega_k(D_2\phi)}{\sqrt{2i\lambda}},\qquad b_{22}=0,
\end{equation*}
such that
\begin{equation*}
b=b_{11}b_{22}-b_{12}b_{21}=-\frac{(\mbox{pr}\omega_k(D_2\phi))^2}{2i\lambda}.
\end{equation*}
For the surface (\ref{F55}), the tangent vectors are linearly independent, so that the immersion defines a surface in $\mathfrak{sl}(2\vert1,\mathbb{G})$.

\section{Conclusions}
In this paper, we have constructed two SUSY versions of the FG formula for the immersion of 2D surfaces in Lie superalgebras. The first (bosonic) SUSY extension is considered for the immersion of a bosonic supermatrix $F$ in the Lie superalgebra $\mathfrak{gl}(p\vert q,\mathbb{G})$ using a bosonic infinitesimal deformation of the LSP and, therefore, of the ZCC. We have provided the form of the tangent vector based on three different deformations, i.e. transformations of the spectral parameter, invariances under gauge transformations of the wavefunction $\Psi$ and symmetries of both the LSP and the ZCC. From the tangent vectors, we have been able to describe the metric using the super Killing form and to find a unit normal vector $N$ which allows us to describe the coefficients of the second fundamental form in terms only of the potential matrices $U_j$ and their deformations $A_j$. The Gaussian and mean curvatures were determined. The second (fermionic) SUSY version of the FG formula for immersion uses an odd-valued parameter instead of the bosonic infinitesimal parameter. This leads to a fermionic supermatrix $F$ immersed in the Lie superalgebra $\mathfrak{gl}(p\vert q,\mathbb{G})$. Using similar deformations as those described in section 4.1 for the bosonic formula for immersion, we have determined the form of the tangent vectors together with a unit normal vector. The two fundamental forms and the two curvatures have also been given explicitly in terms of the potential matrices $U_j$ and their deformations $A_j$ via the super Killing form. 

The integrable SUSY sine-Gordon equation and its Lax pair have been employed in order to apply the two SUSY versions of the FG formula for immersion. Among these examples, we have considered a bosonic deformation of the spectral parameter for which the Gaussian and mean curvatures resemble the classical case. Moreover, the Gaussian curvature is constant and positive. We have also considered separately the bosonic and fermionic gauge transformations and provided their associated geometric characterizations. Both surfaces are nonlinear Weingarten-type surfaces.  The bosonic variable translations and the SUSY transformation symmetries have also been investigated.

This research could be extended in several directions. It would be interesting to investigate other examples of integrable SUSY systems like the SUSY Schr\"odinger equations or the SUSY Korteweg--de Vries equation and the associated soliton surfaces. Also, the use of different norms and inner products could be applied to get a different approach depending on the physical interpretation of the considered models. Moreover, it would be interesting to explicitly solve the wavefunction $\Psi$ so that we can provide a visual image of the surface. As an additional future perspective, we could investigate how the conserved quantities, such as the Hamiltonian structure, manifest themselves on the surface.

\section*{Acknowledgements}
AMG's work was supported by a research grant from NSERC. SB acknowledges a doctoral fellowship provided by the FQRNT of the Gouvernement du Qu\'ebec.

\end{document}